# Exciton, trion and localized exciton in monolayer Tungsten Disulfide

Aïda Hichri,[1,a)] Imen Ben Amara,[2] Sabrine Ayari[1] and Sihem Jaziri[1,2]

[1]*Laboratoire de Physique des Matériaux, Faculté des Sciences de Bizerte 7021 Jarzouna, Université de Carthage, Tunisie.*
[2]*Laboratoire de Physique de la Matière Condensée, Département de Physique, Faculté des Sciences de Tunis, Université Tunis el Manar, Campus Universitaire 1060 Tunis, Tunisie.*

The ultrathin transition metal dichalcogenides (TMDs) have emerged as promising materials for various applications using two dimensional (2D) semiconductors. They have attracted increasing attention due to their unique optical properties originate from neutral and charged excitons. Here, we report negatively charged exciton formation in monolayer TMDs, notably tungsten disulfide $WS_2$. Our theory is based on an effective mass model of neutral and charged excitons, parameterized by ab-initio calculations. Taking into the account the strong correlation between the monolayer $WS_2$ and the surrounding dielectric environment, our theoretical results are in good agreement with one-photon photoluminescence (PL) and reflectivity measurements. We also show that the exciton state with *p*-symmetry, experimentally observed by two-photon PL emission, is energetically below the 2*s*-state. We use the equilibrium mass action law, to quantify the relative weight of exciton and trion PL. We show that exciton and trion emission can be tuned and controlled by external parameters like temperature, pumping and injection electrons. Finally, in comparison with experimental measurements, we show that exciton emission in monolayer tungsten dichalcogenides is substantially reduced. This feature suggests that free exciton can be trapped in disordered potential wells to form a localized exciton and therefore offers a route toward novel optical properties.

aida.ezzeddini@gmail.com

**I. INTRODUCTION**

Research exploring 2D layered materials, such as graphene, hexagonal-boron nitride,[1] TMDs[2-6] and black phosphorus,[7,8] are attracting increased focus for many applications in electronics, photonics and optoelectronics. Single layer of TMD materials has been a favored subject (like graphene, TMDs have hexagonal crystal structure composed of layers of a transition metal M from group VI sandwiched between two layers of chalcogen atoms X with a generalized formula of $MX_2$). Bulk crystal, in which the layers interact via van der Waals forces, is general an indirect gap semiconductor. With the reduction in number of layers, the indirect band gap increases and transforms to direct band gap for a monolayer TMDs[9-11] such as $WS_2$, $WSe_2$, $MoS_2$ and $MoSe_2$. This class of semiconducting TMDs is particularly interesting for optoelectronic applications. Recently, they have been used for non-linear optics and optoelectronic such as photodetectors,[12] light emitting devices[13] and solar cells.[14] In addition to the semiconducting behavior, TMDs exhibit diverse properties that depend on their composition and can be semimetals (e.g., $WTe_2$, $TiSe_2$), metals (e.g., $NbS_2$, $VSe_2$), and superconductors (e.g., $NbSe_2$, $TaS_2$). The chemistry of $MX_2$ compounds offers new opportunities for going beyond graphene. Experimental and theoretical studies show that PL in monolayer TMDs can be many orders of magnitude stronger than that in their bulk counterpart. Due to the strong quantum confinement, reduced dielectric screening and large carrier effective masses, Coulomb interaction is dramatically enhanced. This leads to large binding energies for both excitons and trions which dominate the optical and electronic properties.[15] Indeed, the observed large exciton binding energies would have a significant impact in the next-generation photonics and optoelectronics applications based on 2D atomic crystals.

The exciton is formed when the electron absorbs a quantity of light, thus it can be moving from the valence band into the conduction band. If the Coulomb-bound electron-hole pair is strong enough when we inject an excess of charges, excitons can capture additional charges to form negatively charged excitons known as trions. The excess charge density can be realized and controlled by different methods like electrical gate[4] and photo-ionized carriers trapped on the donors at low temperature .[16] In some TMDs, the tunability of the excitonic species can also be established by chemical doping through the absorption of gases and organic molecules.[17,18] The investigation of the trion in TMDs is accomplished by several theoretical and experimental studies[16,18-25] such as scanning tunneling spectroscopy, temperature-dependent PL and nonlinear optical spectroscopy. The obtained binding energies of trions are about 18-30 meV in monolayer $MoS_2$,[20] $MoSe_2$[4] and $WSe_2$ [13] and those of excitons in monolayer $MoS_2$,[2,20,26] $MoSe_2$,[4,27] $WSe_2$ [3,13,15,28] and $WS_2$ [15,16,18,23,29-31] are in a range of 200 to 800 meV. Actually, the exact values are still the subject of discussion, especially for monolayer $WS_2$ where the binding energy of the exciton or/and trion is still controversial.[14-16,19] In comparison with the 2D standard semiconductor quantum wells[32,33] where the trion binding energy is only a few meV, TMDs are characterized by a large magnitude of trion binding energy, which has never been observed except for monolayer phosphorene.[34] Besides, the stability of excitons and trions in monolayer TMDs has been observed even at room temperature[20,35] in contrast to the black phosphorus that suffers of rapid oxidation in ambient conditions.[36]

The difference between monolayer tungsten and molybdenum dichalcogenides is the size of the spin-orbit splitting due to the difference size of the transition metal atom. Therefore, spin-orbit splitting between the A and B excitons is approximately 400 meV for both $WS_2$ and $WSe_2$.[37] These excitons labeled A and B, correspond to the lowest energy exciton states originated from transitions from the two highest energy spin orbit split-off valence bands to the lowest energy conduction bands around the K (K') point in the Brillouin zone. Moreover, the electronic transitions in the $WS_2$ samples present tight spectral features, authorizing identification and analysis of the excited states and the possibility of the comparison with theoretical results.[30] In this paper we will focus only on the A exciton series in monolayer $WS_2$.

Generally, electroluminescence or PL intensity is characterized by excitonic emission when the monolayer is optically pumped. Increasing the carrier density, trion emission becomes important due to the enhancement of trion formation. It follows that the change of the carrier concentration strongly influences optical properties, in particular PL measurement. Large PL intensity was also observed from induced defects[38] which can be created by different mechanisms, like alpha-particle irradiation,[39] residual impurities,[40] atomic vacancy formation,[41] strain from the substrate, high-temperature annealing, impurity doping[42] and chemical functionalization.[41] Especially, disorder derived from defects can lead to a dramatic change in the physical behavior of the interband excitations, producing an inhomogeneous spectral broadening and localization. Experimental measurements[43] show that the defect peak is located below the free exciton and trion emission this means that defects act as very efficient exciton traps, reduce the electron-hole mobility and therefore create localized states.

In this work, we use a theoretical framework that takes into account the substrate effect caused by the strong dielectric screening, to study exciton and trion binding energies. In the law density limit, we will be interested in trion formation as a function of temperature and injection electrons. Finally, we will show that disorder potential arising from defects strongly affects the exciton states and leading to inhomogeneous broadening of the exciton resonance. To offer the required parameters for the applied theoretical model, we carried out the ab-initio calculations by means of Density Functional Theory (DFT).[44] The optoelectronic response of $WS_2$ monolayer, including the band structure, effective masses of charges carriers, densities of states and dielectric constants, is therefore determined.

**II. Ab- INITIO CALCULATIONS:** Numerical details and results

Based on self-consistent scheme, the present first principle study consists on solving the Kohn–Sham (KS) equations [45] by using the all electron Full-Potential Linearized Augmented Wave plus local orbital (FP-LAPW+lo)[46] method as embedded in WIEN2k simulation package.[47] In this approach, the wave functions are expanded as combination of spherical harmonic functions up to $l_{max} = 10$ inside the muffin tin (MT) spheres surrounding the atomic site. Whereas in interstitial regions of the unit cell, the plane wave with a cut-off of $R_{MT}K_{max} = 7$ is used. $R_{MT}$ and $K_{max}$ design the muffin-tin radii and the maximum modulus for the reciprocal lattice vectors, respectively. The charge density is Fourier series expanded up to $G_{max} = 12$. Values of $R_{MT}$ are selected to ensure the core density confinement in the non-overlapping MT spheres. We have employed the new generalized gradient approximation as presented by Engel and Vesko (EV) in order to reproduce better the exchange-correlation[48] potential compared to others formalism such as LDA [49] and GGA.[50]

Accurate Irreducible Brillouin zone (IBZ) integrations are performed using the modified tetrahedron method of Bloch et al[51] with a denser mesh of 150 special k-points. The monolayer $WS_2$ is modeled by constructing the 1×1×1 supercell with a vacuum thicknesses in the z-direction chosen around 35 Bohr in the aim to decouple the periodically repeated systems. The electronic structure calculations are converged in the accuracy of $0.1 mRy$ in the total energy of the herein material.

The optical features are determined in terms of frequency complex dielectric function $\varepsilon(\omega) = \varepsilon_1(\omega) + i\varepsilon_2(\omega)$, with small wave vector, in the tensor representation. The investigated compound with hexagonal symmetry has two non-zero diagonal components of the dielectric tensor. These components correspond to an electric field perpendicular and parallel to the z-axis, namely, $\varepsilon_1^{xy}(\varepsilon_1^{\perp})$ and $\varepsilon_1^{zz}(\varepsilon_1^{\parallel})$. The imaginary dielectric function, or the absorptive part of the dielectric function, is presented as a superposition of direct and indirect transitions. However, the indirect ones, which involve the scattering of phonon, are neglected due to the small contribution to $\varepsilon_2$. Hence, the components of the imaginary $\varepsilon_2$ are calculated by summing all possible direct interband transitions between the occupied and empty state using the relation given in Ref [52].The real dielectric function $\varepsilon_1$ is therefore given by means of the Kramer Kronig transformation.[53]

To identify promising materials for optoelectronic, the knowledge of band gap is crucial. Hence, we have calculated the band structure of WS$_2$ monolayer along some high symmetry k-points of the first BZ using the efficient EVGGA approach for exchange-correlation potential treatment. Fig.1a which illustrates the energy bands together with the total densities of states (DOS) spectrum indicates that WS$_2$ monolayer is K-direct gap semiconductor with band gap of 2.24 eV. Our elucidated parameter is compared with the available previous theoretical [54,55] and measured ones[56] in the Table, and it seems to be in good agreement with the experimental results.[56] This highlights the effect of the used EVGGA which yields a better bands splitting. The electron and hole effective masses ($m_e, m_h$) of WS$_2$ monolayer are evaluated in terms of free electron effective mass $m_0$ and obtained from the conductive and valence band curvature by fitting parabolas at the K point from Figure 1(a). The calculated component $\varepsilon_1^{\perp}(\omega)$ and $\varepsilon_1^{\parallel}(\omega)$ versus incident photon energy are illustrated in Figure 1(b). The plot shows different thresholds peaks. Indeed, the most important ones are the zero frequency limits, i.e. the static dielectric constant $\varepsilon_1^{\perp}(0)$ and the maximum value in the plan *xy*. However, for our analytical model, we will only focus on the estimated maximum value which is mentioned in the Table.

### III. FREE EXCITON AND TRION

To estimate the binding energies and the oscillator strength of the exciton and trion states in monolayer WS$_2$, we use 2D effective-mass approximation based on our first principle calculations. It is convenient to work in the center-of-mass frame, the physics of exciton is described by the following Hamiltonian $H_X = H_{CM} - \frac{\hbar^2 \nabla_\rho^2}{2\mu} + V_{2D}(\rho)$, where $H_{CM}$ is the center of mass Hamiltonian, $\mu = \frac{m_e m_h}{m_e + m_h}$ is the reduced effective mass in the 2D ($x$, $y$) plan, and $V_{2D}$ is a nonlocally-screened electron-hole interaction due to the screening caused by the change in the dielectric environment.[57] Using the polar coordinates, the in-plane 2D dielectric screened Coulomb interaction for electrons and holes located respectively at the position $\boldsymbol{r_e}$ and $\boldsymbol{r_h}$ and separated by the relative coordinates $\boldsymbol{\rho} = \boldsymbol{r_e} - \boldsymbol{r_h}$ has the following expression $V_{2D}(\rho) = -\frac{e^2}{\varepsilon} \int \frac{q_s dq}{q + q_s} J_0(q\rho)$, $e$ is the electron charge, $J_0(X)$ is the Bessel function of the first kind, $\varepsilon$ is the relative dielectric constant given by $\frac{\varepsilon_1 + \varepsilon_2}{2}$ and $q$ is the electron's wave vector. The screening 2D parameter is $q_s = \frac{1}{r_s}$, where $r_s = \frac{\epsilon_r d}{\varepsilon_1 + \varepsilon_2}$ is the screening radius which gives a crossover length scale between the long and small range Coulomb's interaction. $q_s$ explains the strong dielectric contrast between the monolayer WS$_2$ and its surroundings. $\varepsilon_1$ (air) and $\varepsilon_2$ (substrate) are the environmental relative dielectric constants, $\epsilon_r$ (see Table) is the WS$_2$ relative dielectric constant and $d = 6.7$Å is the effective width of the WS$_2$ layer. The free motion of the center of mass is characterized by the Hamiltonian $H_{CM} = -\frac{\hbar^2 \nabla_R^2}{2M}$, where $\boldsymbol{R} = \frac{m_e \boldsymbol{r_e} + m_h \boldsymbol{r_h}}{m_e + m_h}$ and $M = m_e + m_h$ are respectively the exciton center of mass coordinate and mass. Here, the in-plane center of mass momentum operator $\boldsymbol{Q}$ is a constant of motion. To solve the eigenvalue equation, we use a wave function expansion technique; it is factorized into: $\Psi(\rho, \theta, \boldsymbol{R}) = \frac{1}{\sqrt{S}} e^{-i\boldsymbol{QR}} \sum_{n,m} c_{nm} \varphi_{n,m}(\rho, \theta)$ where $S$ is the area of the monolayer. The wave functions $\varphi_{n,m}(\rho, \theta)$ are the 2D-hydrogenic states; It expanded in terms of orthogonal associated Laguerre polynomials $L_{n'}^\alpha(x)$ and it's given by: $\varphi_{n,m}(\rho, \theta) = $

$$e^{im\theta} a_{n,m} \left(\frac{4\rho}{(2n-1)a_B}\right)^{|m|} e^{-\frac{2\rho}{(2n-1)a_B}} L_{n-|m|-1}^{2|m|}\left(\frac{4\rho}{(2n-1)a_B}\right), \quad a_{n,m} = \frac{1}{\sqrt{2\pi}}\left\{\left(\frac{2}{(2n-1)a_B}\right)^2 \frac{(n-|m|-1)!}{(n+|m|-1)!(2n-1)}\right\}^{1/2}$$

$n = 1, 2, 3 ...$ with angular momentum $m = 0, \pm 1, \pm 2, \pm 3, ... \pm n - 1$. The states are (2n-1) fold degenerate, the states are labeled *s* for *m*=0, *p* for *m*=$\pm 1$ and *d* for *m*=$\pm 2$. In order to compare our results to experimental findings, we have limited ourselves to the *5s* and *3p* states. In this work, we use as unit of length, the three dimensional Bohr radius $a_B = \frac{\varepsilon \hbar^2}{\mu e^2}$ for which the 2D Bohr radius is $a_{2D} = \frac{a_B}{2}$. For the relative dielectric constant $\varepsilon_r = 5.8$ and when the WS$_2$ layer is on the top of SiO$_2$ (300nm)/Si substrate characterized by $\varepsilon_2 = 3.9$ while the top surface is exposed to air ($\varepsilon_1 = 1$), the screening length $r_s$ is equal to 8 Å. The relative energy levels $E_{(rel)\,nm}$ and corresponding eigenvectors $\psi_{nm}$ are obtained numerically by direct diagonalization of the full matrix resulting from the projection of the Hamiltonian $H_X$. Indeed, the $n^{th}$ *s* (*p*) exciton transition energy is given by $E_{nm} = E_g + E_{(rel)\,nm} + E_{CM}$, where $E_g$ is the band gap and $E_{CM} = \frac{\hbar^2 Q^2}{2M}$ is the center of mass energy.

We report in Figure 2 (a) the first positions of the excitonic binding energy $B_{nm} = -E_{(rel)\,nm}$ obtained by solving the Hamiltonian $H_X$ (black squares) and results extracted from experimental data. The states labeled as *1s*, *2s* and *3s* (blue circles) are observed experimentally by linear reflection.[30] *2p* and *3p* states (red stars) are not optically bright in one-photon spectra, but can be analyzed using two-photon excitation spectroscopy.[31] The *np* states can also be detected via resonances in a second harmonic generation. The $E_{2p}$ and $E_{3p}$ states observed in our results are nearly conform to experimental observations.[31] This implies that the screened Coulomb potential in a 2D system given by L.V. Keldysh and plotted in the inset of Figure 2 (a), can explain the behavior of exciton in the most of the 2D TMDs. Note that the potential is used for a system consisting of thin dielectric sheet and it is only valid when the dielectric constant of the sheet is larger than the dielectric constant of the surrounding medium $\left(\varepsilon_r > \frac{\varepsilon_1+\varepsilon_2}{2}\right)$. These interaction potential is compared in Figure 2 (b) with the average dielectric constant of the environment varying from 1 to 10. This plot show strong screening by the 2D sheet at short distances and approaching Coulomb interaction at $r_s \leq \rho < d$. However, in thin films the interaction between charges increases with decreasing thickness $d$, since for $\rho > d$ the field produced by these charges in the environment surrounding the monolayer begins to play a perceptible role.[57] Finally, the large binding energies, non-hydrogenic spectra, and the sensitivity of the exciton binding energies to substrate screening are all found to be a consequence of the broad dispersion of results, that further work is necessary to fully understand these novel quasi-2D materials.

The strong Coulomb interaction in monolayer TMDs leads to the formation of tightly bound trions, which can be seen generally at low temperatures. We adopt the same formalism to investigate the trion binding energy in monolayer $WS_2$. The trion Hamiltonian in the effective mass approximation with screened interactions is given by: $H_T = \sum_i H_{Xi} - \frac{1}{2m_h}\boldsymbol{\nabla}_{\rho 1}\boldsymbol{\nabla}_{\rho 2} - V_{2D}(\boldsymbol{\rho}_1 - \boldsymbol{\rho}_2)$, $i$=1, 2 refer to the first and the second exciton corresponding to the relative coordinates $\boldsymbol{\rho}_1 = \boldsymbol{r}_{e1} - \boldsymbol{r}_h$ and $\boldsymbol{\rho}_2 = \boldsymbol{r}_{e2} - \boldsymbol{r}_h$. We notice that the three particles bound state of negatively or positively charged excitons are analogous because electrons and holes in $WS_2$ have nearly the same effective mass. We focus on the study of a spin singlet charged exciton A, related to the fundamental band gap of the material (bound states for triplet trions have not been observed neither in experiment[58] nor in theory.[59] We use relative $(\boldsymbol{\rho}_1, \boldsymbol{\rho}_2)$ and center of mass $\boldsymbol{R}_T = \frac{M\boldsymbol{R} + m_e \boldsymbol{r}_{e2}}{M_T}$ coordinates of the two quasi-particles (exciton 1 and exciton 2) with $M_T = 2m_e + m_h$ is the trion mass and $\boldsymbol{Q}_T$ is the trion's center of mass wave vector. Since the in-plane trion center-of-mass motion is also separated from the relative motion, then the trion wave function can be written as the product of the trion center of mass contribution and the relative wave function : $\chi_{n,m,n',m'}(\boldsymbol{\rho}_1, \boldsymbol{\rho}_2, \boldsymbol{R}_T) = \frac{1}{\sqrt{S}} e^{-i\boldsymbol{Q}_T \boldsymbol{R}_T} \varphi_{n,m}(\boldsymbol{\rho}_1) \times \varphi_{n',m'}(\boldsymbol{\rho}_2)$. Eigenvalues $E_{nT}$ and eigenvectors $\psi_{nT}$ are obtained by numerical diagonalization. Having the trion wave function at our disposal, we can calculate the eventual optical transitions.

We outline the optical properties of the three-particle bound state of 1$s$ exciton with an additional electron, as shown in the one-photon PL spectra.[16] The oscillator strength is proportional to the probability of finding one of the two electrons and hole in the same unit cell i.e. the square of the exciton 1 or exciton 2 relative motion wave function $\psi_{1T}(\boldsymbol{\rho_1}, \boldsymbol{\rho_2}, \boldsymbol{R}_T)$ for $\boldsymbol{\rho_1} = 0$ (or $\boldsymbol{\rho_2} = 0$). Using the Fermi's golden rule, we calculate the oscillator strength given by $F(\boldsymbol{Q}_T) = \left| \int d\boldsymbol{\rho}_2 \psi_{1T}(0, \boldsymbol{\rho}_2) \exp\left(-\frac{i\boldsymbol{Q}_T \boldsymbol{\rho}_2 M}{M_T}\right) \right|^2$ where $\psi_{1T}(0, \boldsymbol{\rho}_2)$ is the trion eigenvector of the lowest excited state. The Boltzmann distribution has been used to model the temperature dependence of the PL in the law density limit; it can be written as follows: $f_c(\boldsymbol{Q}_T) = \frac{\hbar^2}{2\pi M_T k_B T} \exp\left(-\frac{\hbar^2 Q_T^2}{2 M_T k_B T}\right)$, T is the temperature and $k_B$ is the Boltzmann constant. The trion one-photon PL lineshape is given by $P^T(\omega) = \int f_c(\boldsymbol{Q}_T) F(\boldsymbol{Q}_T) \delta\left(\hbar\omega - E_{1T} + \frac{\hbar^2 Q_T^2}{2m_e} \frac{M}{M_T}\right) d\boldsymbol{Q}_T$, where $\hbar\omega$ is the emitted photon energy. We use Lorentzian curves to fit the measured PL spectra. Note that the exciton PL lineshape has the following expression $P^X(\omega) = S|\psi_{1s}(\boldsymbol{\rho} = 0)|^2 \int f_c^X(\boldsymbol{Q}) \delta\left(\hbar\omega - E_{1s} - \frac{\hbar^2 Q^2}{2M}\right) d\boldsymbol{Q}$. The exciton density is distributed according to the Boltzmann distribution $f_c^X(\boldsymbol{Q}) = \frac{\hbar^2}{2\pi M k_B T} \exp\left(-\frac{\hbar^2 Q^2}{2M k_B T}\right)$. In order to compare with experimental data,[16] we plot in Figure 3(a) the exciton and trion one-photon PL emission calculated at T=4K. Using our parameters, we observe a neutral exciton emission at $\sim 1.987 \, eV$ and trion emission at $\sim 1.951 \, eV$ yielding a large trion binding energy of $\sim 35 \, meV$. Notice that trion binding energy is conventionally introduced as $B_T = E_g + E_{1T} - E_{1s}$. Figure 3 (b) illustrates a color map of the PL spectrum of trion in monolayer WS$_2$ as a function of temperature. There is a pronounced decrease of PL intensity when increasing temperature. To interpret the tuning of the ratio between trion and exciton emissions, given in ref [16], we will show in the next section that the exciton and trion PL intensities can be modeled by the mass action law.

For the quantification of the relative weight of exciton and trion PL, we apply an equilibrium mass action law between excitons, trions, free electrons and free holes. This discuss the dependence of the emission on temperature, background electron density $n^B$ and the pumping laser excitation density $n^P$. The relation between these populations, at a given temperature T, can be estimated as

$$\frac{n_X n_e}{n_T} = \frac{2m_e}{\pi \hbar^2} \frac{M}{M_T} k_B T \times exp\left(-\frac{B_T}{k_B T}\right) = A \quad (1a)$$

$$\frac{n_h n_e}{n_X} = \frac{m_h}{2\pi \hbar^2} \frac{m_e}{M} k_B T \times exp\left(-\frac{B_{1s}}{k_B T}\right) = B \quad (1b)$$

Note that the determination of the trion concentration $n_T$ with binding energy $B_T$, requires the knowledge of both exciton $n_X$ and electron $n_e$ densities. We adopt a steady state model of all the particles in our system, to determine the intensity of PL signals. By using a suitable pumping density $n^P$ we create photoexcited electrons to increase the neutral exciton emission. The number of photoexcited electrons is $n^p = n_h + n_X + n_T$, where $n_h$ is a free hole density. Laser power above band gap excitation, in addition to create electron-hole pairs, can photo-ionize the carriers trapped on the donors creating an excess electron density in conduction band such as: $n^B = n_e + n_T - n_h$. $n^B$ is the background electron density before subband gap laser excitation. $n^B$ and $n^p$ are external parameters that can be controlled by laser intensity or by gate voltage, whereas $n_X, n_T, n_e$ and $n_h$ are steady state variables. From charge conservation, we write

$$n^p + n^B = n_e + n_X + 2n_T \qquad (2)$$

By solving these equations, we obtain $n_e^3 + \beta n_e^2 + \gamma n_e + \delta = 0$, where $\beta = A + n^p - n^B$, $\gamma = A(B - n^B)$ and $\delta = -AB(n^P + n^B)$. In order to simplify the resolution of the third degree equation, we start by evaluating the order of magnitude of our parameters. Since the exponential term in Eq. (1.b) decays rapidly to zero due to the large value of exciton binding energy (~260meV) compared to that of trion (~35meV), we, therefore, neglect the free hole density in the system.

In Figure 4(a) and Figure 4(b) we show the trion (red circles) and exciton (black squares) densities as a function of temperature and pump to background ratio $s = n^p/n^B$. The plot shows that the maximum of trion density is nearly equal to the saturated exciton PL when $n_T$ vanishes. This behavior is derived from the conservation of the total number of exciton and trion with increasing temperature or decreasing background density. With the increasing of temperature the $n_T$ signal decreases significantly around 100K, this is due to the rupture of the bound trion state due to thermal fluctuations. The evolution of the densities $n_X$ and $n_T$ illustrated respectively in Figure 4(c) and Figure 4(d) as a function of temperature and *s* for a fixed $n^P$, demonstrates the ability of temperature, doping and injection to tune $n_X$ and $n_T$. By plotting these quantities as a function of *T* and *s*, we are able to model the full range of experimental data. The formation process of trions[60] which is promoted by the excess of negatively charged carriers is strongly temperature dependent. Applying the mass action law with a specific trion binding energy, one can get an idea about the amount of excess electron density. Since, according to the experimental spectra of flake 1 in the ref. [16], we plot in Figure 5 the ratio between exciton and trion densities as a function of temperature. The rate is shown for different values of the pump-to-background ratio *s*. The cross symbols are values which have been deduced from the measured PL spectra by the above mentioned experiment. The symbols fall on different curves. At low temperature $T \lesssim 100K$, we notice that for $n^P = 2.10^{10} cm^{-2}$ and *s*=0.5 or 1 the rate varies slowly as a function of *T*. This feature is consistent with previously reported results, where the trion formation has been reported to be dominant with doping and low temperature. Whereas, with decreasing $n^P$, the PL spectrum is dominated by the exciton peak, which strongly suggests that the excitons can recombine without forming trions because of the decrease in the number of excess carries in the monolayer. However for the following temperatures T=30, 70, 110, 125K dictated by the experiment,[16] we expected to have a much larger pumped density for the higher temperature. The reasonable well agreement of the experimental exciton and trion PL line-shape with our calculations based on the mass-action low leads to the conclusion that for each temperature we can know approximately the value of the background electron and pumping densities. Since, the trion formation can be modulated by controlling the carrier density and temperature.

Recently, due to the quality of samples, doping and measurement conditions, the studies on trion and exciton dynamics[61-71] have shown considerable discrepancy in the interpretation of the experimental results. Its signature is a large band appearing a few meV below the trion and exciton emission at low temperature. In the follow, we assigned this band as localized states due to the trap of free exciton in a disordered potential.

**IV-LOCALIZED EXCITONS**

In recent years, defect-derived excitonic states have received considerable attention for their potential applications in single-photon emitters.[71, 72] Points defects can trap free charge carriers and create localize states as it is reported by some groups.[17,67,73,74] At low temperature, excitation power dependence measurements proved the existence of multiple emission peaks in monolayer TMDs flakes.[43,74] In WSe$_2$, localized exciton that has been investigated through single photon emission experiments,[71] is red-shifted tens of meV below the trion transition. The spectrum shows five pronounced PL peaks, which are observed at 10K. In monolayer WS$_2$ the spectral of excitons bound to defects that were created by plasma treatment, is located at 100 meV below the A-exciton emission peak.[39] On the other hand, monolayer Nb-doped $WS_2$ shows news PL peaks between 1.4-1.6eV below the free exciton which is located at approximately 2eV.[42] These emission peaks located below the free-exciton and trion energies suggests the presence of an effective exciton "mobility edge", i.e., below (above) a certain energy, the center-of-mass motion of the excitons is localized (delocalized).[67] Then, in the presence of defect, the momentum of the center-of-mass motion is no longer a good quantum number as was assumed in the previous model related to the free exciton. Such effects introduce an additional confinement in all directions like the case of a quantum dot potential. However, the exciton in monolayer TMDs behaves as a massive particle subject to a disordered potential, leading to spatially localized eigenstates of the center of mass motion. In order to prove further insight into the behavior of the defects states, we use a theoretical model which allows us to reproduce the multiple emission peaks observed in monolayer tungsten dichalcogenides. As mentioned below, we interest only on the influence of localization on the $1s$ states. We will suppose that trapping by deep or shallow defect well can localize exciton and create a multiple emission peaks.

In the beginning, we consider that the previous free exciton is trapped in local potential wells that can be modeled simply in the form of a Gaussian well. Due to its continuity and finite depth and range, the Gaussian potential fairly well approximates the real confinement.[75] Taking account the center of mass and relative coordinates, we write the Hamiltonian of the localized states of the exciton as follows:

$$H^{loc} = H_{rel} + H_{cm} + V\left(\mathbf{R} + \frac{m_e\boldsymbol{\rho}}{M}\right) + V\left(\mathbf{R} - \frac{m_h\boldsymbol{\rho}}{M}\right) \tag{3}$$

where $V\left(\mathbf{R} \pm \frac{m_{e(h)}\boldsymbol{\rho}}{M}\right) = -V_0 \, exp - \frac{1}{2R_0^2}\left(\mathbf{R} \pm \frac{m_{e(h)}\boldsymbol{\rho}}{M}\right)^2$ is the Gaussian potential applying separately on the electron and hole respectively. $V_0$ is the height of the potential well and $R_0$ is the range of the confinement potential, which corresponds to a radius of the zero-dimensional $WS_2$ flake. According to the band structure given by DFT calculation, we can get $m_e = m_h$ for the monolayer $WS_2$ as is seen in Figure 1 (a). On the other hand, the sample size is in the range of 100 $nm$, then $\left|\frac{\mathbf{R} \pm \frac{\boldsymbol{\rho} m_{e(h)}}{M}}{R_0}\right| \leq 1$ and the Gaussian potential can be approximated to the parabolic potential. Applying this approximation, localized exciton Hamiltonian can be rewritten as follows $H^{loc} = -\frac{\hbar^2 \boldsymbol{\nabla}_\rho^2}{2\mu} + V_{2D}(\rho) + V_0 \frac{\rho^2}{4R_0^2} - \frac{\hbar^2 \boldsymbol{\nabla}_R^2}{2M} + V(R)$. According to this equation, an additional term $V_0 \frac{\rho^2}{4R_0^2}$ appears which represents a weak parabolic confinement that affects the relative exciton motion. This parabolic potential is dominated by the screening potential $V_{2D}(\rho)$ because when we evaluate the matrix element of the 1s state, respectively for the 2D parabolic potential and the non-local screening we obtain the following values $\langle \varphi_{1s}|V_{2D}(\rho)|\varphi_{1s}\rangle = -0.9 eV$ and $\frac{V_0}{4R_0^2}\langle \varphi_{1s}|\rho^2|\varphi_{1s}\rangle = 1.42\ 10^{-5} eV$. This estimation allows as to expanded the wave functions of the relative part of $H^{loc}$ in terms of the hydrogenic wave functions $\varphi_{n,m}(\rho,\theta)$. Besides, the center of mass motion is no longer free, as it was for the delocalized exciton; their corresponding eigenstates are built using Hermite polynomials. Therefore, the total wave function can be written as $\Psi_{\tilde{n}_X,\tilde{n}_Y,\tilde{n},\tilde{m}}(R,\rho,\theta) = \sum_{n_X,n_Y,n,m} c_{n_X,n_Y,n,m}\ \phi_{n_X,n_Y}(R)\varphi_{n,m}(\rho,\theta)$ where $\phi_{n_X,n_Y}(R)$ is a 2D harmonic oscillator state with a frequency $\omega_{cm} = \sqrt{\frac{2V_0}{MR_0^2}}$ and an energy $E_{cm} = \hbar\omega_{cm}(n_X + n_Y + 1)$, $n_X$ and $n_Y$ are positive integers or zero. The $H^{loc}$ matrix is numerically diagonalized to produce the localized exciton eigenvalues and associated eigenvectors.

To explore the effect of the three-dimensional quantum confinement of excitons in monolayer TMDs, we calculate the PL emission. Our numerical computation is carried out for WS$_2$ monolayer. In order to well reproduce the experimental findings, we start by varying the Gaussian potential deep to extract the adequate parameters. Figure 6(a) shows the exciton center of mass energy confined in a respectively shallow and intermediate Gaussian potential with a large radius around 25x the Bohr radius and for different well deep about 0.1 $R_y$, 0.27 $R_y$ and 0.5 $R_y$ where $R_y = \frac{e^4\mu}{2\varepsilon^2\hbar^2}$ is the three-dimensional exciton Rydberg constant. According to experimental measurement, the localized exciton states are located at tens of meV from free exciton position. The depth of the potential well, i.e., the confinement energy, is given by the spacing between the PL emission energy of the localized exciton and the free exciton emission Therefore, it is preferable to choose the second case i.e. a potential-well depth $V_0 = 0.27\ R_y$ for which $R_0 = 25\ a_B$. As is seen in this figure, the Gaussian potential reasonably well approximates the confinement of the center of mass exciton motion. The calculated PL spectrum is illustrated in Figure 6 (b), we respectively refer to free and the localized excitonic states as X and LX. This spectrum is significantly different from the spectral characteristics of pristine WS$_2$ monolayer where only two peaks were observed at ~1.95 and ~1.98eV. However, the PL intensity of the free exciton X is relatively reduced in the presence of the localized exciton. We assume that this attenuation is attributed to the trap of free neutral excitons into the defect potential. Additionally, five pronounced PL peaks from WSe$_2$ flak are also shown in the inset. According to the ref [42,74], defect-related localized state transitions is the most likely origin of these peaks. The free exciton in WSe$_2$ monolayer is located at $\approx 1.75$eV,[43] then the depth of the potential well is taken around 130meV and $R_0 = 100$Å. The band gap and the effective masse of the charge carriers are given in the table. Other experimental measurement relative to the monolayer WS$_2$,[42] prove the existence of multiple PL peaks placed in a wide range of energy (1.4-1.6 eV). To modelize these experimental results, it is necessary to use a deep Gaussian potential trap in which free exciton is strongly spatially localized.

The confinement energies and the spacing of the energy levels are smaller compared to the binding energy and the spacing of the energy levels of the delocalized exciton. Notice that the energy difference between the ground and the first excited states of delocalized exciton in monolayer WS$_2$ (around 180 *meV*) is larger or comparable to the depth of the confinement potential of the localized excitons which is about 100 *meV*. This feature is in contrast to topical semiconductor quantum dot systems, where the confinement potential is much larger than the exciton binding energy. However, the remarkably feature of our results is the equidistance between the localized states. This situation contradicts the experimental finding where the sharp emission peaks originate from localized excitons is characterized by an anharmonic spectrum. However, the localized exciton energies vary due to different local potentials arising from disorder and impurities, which inhomogeneously broaden the optical linewidth. In the following, we assume that inhomogeneity and anharmonicity will be related to disorder-induced band-tail states in monolayer tungsten dichalcogenide.[70] Because the perturbation introduced by disorder is not sufficient to produce a transition from the exciton 1$s$ state to higher sates of the relative exciton motion $(E_{2s} - E_{1s}) \approx 0{,}18eV$, so only its center-of-mass motion is affected by disorder.[76, 77]

The reliable model for the effective center of mass potential *V(X,Y)* is given in terms of a Gauss-distributed spatially-correlated random potential[78] characterized by zero-mean $\langle V(X,Y) \rangle = 0$ and variance given by $\sigma_v^2 = \langle (V(X,Y) - \langle V(X,Y) \rangle)^2 \rangle = \frac{V_0^2}{2}$. To achieve this, the potential function *V(X,Y)* is given by[79]

$$V(X,Y) = \sqrt{\frac{2}{N}} V_0 \sum_j^N \cos(\frac{2\Pi}{L} X\cos\theta_j + \frac{2\pi}{L} Y\sin\theta_j + \beta_j) \tag{4}$$

This potential is created from a superposition of N random plane waves with random direction $\theta_j$, random phase $\beta_j$ (these two parameters are uniform distributed on $[0,2\pi]$), a correlation localized length L and the depth $V_0$. Denote that if N is sufficiently large then $V(X,Y)$ is a large sum of random variables and by the center limit theorem, $V(X,Y)$ is a Gaussian. In the follow, we will use a Gauss correlated disorder potential which emphasizes more the random island structure.

Owing to the experimental data, we consider adequate parameters to reproduce the inhomogeneously broadened localized exciton emission. In Figure 7(a), we show the lakes-and-hills inhomogeneous disorder potential landscape. The simulation parameters are $V_0 = 110 meV$, N=500 and the length scale of the monolayer islands is fixed for $L = 45 Å$. The likes constitute the trap potential which confine the exciton center of mass motion. According to this profile, we plot in Figure 7(b) the corresponding localized exciton PL in monolayer $WS_2$ calculated as follows:

$$P^{LX}(\omega) \propto |\psi_{1s}(\boldsymbol{\rho} = 0)|^2 \left|\sum_{n_X,n_Y} c_{n_X,n_Y} \int \phi_{n_X,n_Y}(\boldsymbol{R}) d\boldsymbol{R}\right|^2 \delta(\hbar\omega - E_{1s} - E_{cm})$$

This spectrum shows a broad emission resonance located around 1.85 and 1.95 eV, the other two energy peaks located at 1.95 and 1.98eV correspond respectively to the trion and free exciton emission. The observed low-energy PL peaks are assigned to emissions from localized excitons trapped in the random potential induced by disorder. Each localized exciton is identified by analyzing the character of the exciton's wavefunction as is seen in Figure 7(c-e). In fact, these states refer to the sequence of quantized state groups in a 2D harmonic oscillator which have the degeneracies 1, 2, 3… These panels display the square modulus of exciton center of mass wave functions selected among the lowest-energy eigenstates. We have chosen to plot in each panel different states having the same features. In these panels, the excitonic states are labeled by an eigenstates number $j$ corresponding to eigenenergy $LX_j$. The states 1, 2, 3 plotted in Figure 7 (c) are attributed to local ground states, (these states are noted state 1 and corresponding to $|\Psi_{0,0,1s}\rangle$). Each of these states is localized in a local minimum of the disorder potential. States with two nodes 4 and 6 corresponding respectively to $|\Psi_{2,0,1s}\rangle$ and $|\Psi_{0,2,1s}\rangle$ (state 2) illustrated in Figure 7 (d) are the first excited states of state 1. We find also in Figure 7 (e) that the states 7, 8, 9 are further excited states of sate 2. According to the experimental considerations, we restrict ourselves only to the ground states because for $j > 3$, $E_j$ are larger than the free exciton energy.

The simulation area used in our calculations is relative to the sizes of the regions probed theoretically. It is known that the poor quality samples with many defects or high impurity density, the PL is characterized by a red shifted emission from the localized states. The quenching of the free excitons emission is a consequence of the exciton relaxation into the localized states. Therefore, disorder acts as exciton traps and reduces the "mobility" of exciton. The PL spectrum of trion, free exciton and localized exciton emission treated by the disorder potential is relatively in the best agreement with experimental finding, indicating that the main features of the localized exciton states are captured by this model. Finally, the impact of localization requires further study because it fundamentally defines the characteristic feature of electronic excitations in the 2D systems.

## V. CONCLUSION

In summary, we have used a modified Wannier-Mott exciton model, in which the significantly reduced dielectric screening of Coulomb interactions is considered, in order to describe the exciton features in 2D TMDs. The required parameters of $WS_2$ monolayer, in particular the gap value, effective mass of the charge carriers and dielectric constants are calculated in the framework of DFT. The presence of strongly bound excitons and trion in monolayer $WS_2$ is directly confirmed from analytical calculations and shown to be in agreement with experimental measurements.[30,31,16] We proved evidence of trion formation with excess background carriers that can be resulting from the doping of the $WS_2$ bulk crystal, with the photo-excited carriers and the photo-ionized carriers trapped on the donors. According to the 2D mass action law, we have shown that lower temperature and relatively higher excitation charge density favorite the formation of trion. These external parameters produce a balance between exciton and trion emission. Finally, we will confront the experimental measurement and by using adequate parameters our theoretical framework is suitable to modelize the disorder in layered tungsten dichalcogenides.

**References**

[1] X. K. Cao, B. Clubine, J. H. Edgar, J. Y. Lin, and H. X. Jiang, Appl. Phys. Lett. **103**, 191106 (2013).

[2] H.-P. Komsa and A. V. Krasheninnikov, Phys. Rev. B **86**, 241201(R) (2012).

[3] K. He, N. Kumar, L. Zhao, Z. Wang, K. F. Mak, H. Zhao, and J. Shan, Phys. Rev. Lett. **113**, 026803 (2014).

[4] J. S. Ross, S. Wu, H. Yu, N. J. Ghimire, A. M. Jones, G. Aivazian, J. Yan, D. G. Mandrus, D. Xiao, W. Yao, and X. Xu, Nat. Commun. **4**, 1474 (2013).

[5] T. Olsen, S. Latini, F. Rasmussen, and K. S. Thygesen, Phys. Rev. Lett. **116**, 056401 (2016).

[6] F. Cadiz, S. Tricard, M. Gay, D. Lagarde, G. Wang, C. Robert, P. Renucci1, B. Urbaszek, and X. Marie, Appl. Phys. Lett. **108**, 251106 (2016).

[7] A. Chaves Tony Low, P. Avouris, D. Çakır, and F. M. Peeters, Phys. Rev. B **91**, 155311 (2015).

[8] J. Yang, R. Xu, J. Pei, Y. W. Myint, F. Wang, Z. Wang, S. Zhang, Z. Yu, and Y. Lu, Science & Applications **4**, e312 (2015).

[9] I. Ben Amara, E. Ben Salem, and S. Jaziri, J. Appl. Phys. **120**, 051707 (2016).

[10] K. F. Mak, C. Lee, J. Hone, J. Shan, and T. F. Heinz, Phys. Rev. Lett. **105**, 136805 (2010).

[11] Q. H. Wang, K. Kalantar-Zadeh, A. Kis, J. N. Coleman, and M. S. Strano, Nat. Nanotechnol. **7**, 699 (2012).

[12] R. S. Sundaram, M. Engel, A. Lombardo, R. Krupke, A. C. Ferrari, P. Avouris, and M. Steiner, Nano. Lett. **13**, 1416 (2013).

[13] J. S. Ross, P. Klement, A. M. Jones, N. J. Ghimire, J. Yan, D. G. Mandrus, T. Taniguchi, K. Watanabe, K. Kitamura, W. Yao, D. H. Cobden, and X. Xu, Nat. Nanotechnol. **9**, 268 (2014).

[14] M. H. Tahersima and V. J. Sorger, Nanotechnol. **26**, 344005 (2015).

[15] A. T. Hanbicki, M. Currie, G. Kioseoglou, A. L. Friedman, and B. T. Jonker, Solid State Commun. **203,** 16 (2015).

[16] A. A. Mitioglu, P. Plochocka, J. N. Jadczak, W. Escoffier, G. L. J. A. Rikken, L. Kulyuk, and D. K. Maude, Phys. Rev. B **88,** 245403 (2013).

[17] S. Tongay, J. Zhou, C. Ataca, J. Liu, J. S. Kang, T. S. Matthews, L. You, J. Li, J. C. Grossman, and J. Wu, Nano. Lett. **13**, 2831 (2013).

[18]N. Peimyoo, W. Yang, J. Shang, X. Shen, Y. Wang, and T. Yu, ACS Nano **8**, 11320 (2014).

[19]T. C. Berkelbach, M. S. Hybertsen, and D. R. Reichman, Phys. Rev. B **88**, 045318 (2013).

[20]K. F. Mak, K. L. He, C. Lee, G. H. Lee, J. Hone, T. F. Heinz, and J. Shan, Nat. Mater. **12**, 207 (2013).

[21]C. Zhang, H. Wang, W. Chan, C. Manolatou, and F. Rana, Phys. Rev. B **89**, 205436 (2014).

[22]Y. Lin, X. Ling, L. Yu, S. Huang, A. L. Hsu, Y.H. Lee, J. Kong, M. S. Dresselhaus, and T. Palacios, Nano Lett. **14**, 5569 (2014).

[23]K. A. Velizhanin, and A. Saxena, Phys. Rev. B **92**, 195305 (2015).

[24]M. Z. Bellus, F. Ceballos, H. Y. Chiu, and H. Zhao, ACS Nano. **9**, 6459 (2015).

[25]J. Shang, X. Shen, C. Cong, N. Peimyoo, B. Cao, M. Eginligil, and T. Yu, ACS Nano **9**, 647 (2015).

[26]Z. Li, Y. Xiao, Y. Gong, Z. Wang, Y. Kang, S. Zu, P. M. Ajayan, P. Nordlander, and Z. Fang, ACS Nano **9**, 10158 (2015).

[27]M. M. Ugeda, A. J. Bradley, S. F. Shi, F. H. da Jornada, Y. Zhang, D. Y. Qiu, S. K. Mo, Z. Hussain, Z. X. Shen, F. Wang, S. G. Louie, and M. F. Crommie, Nat. Mater. **13**, 1091 (2014).

[28]G.Wang, X. Marie, I. Gerber, T. Amand, D. Lagarde, L. Bouet, M. Vidal, A. Balocchi, and B. Urbaszek, Phys. Rev. Lett.**114**, 097403 (2014).

[29]I. Kylänpää and H-P Komsa, Phys. Rev. B **92,** 205418 (2015).

[30]A. Chernikov, T. C. Berkelbach, H. M. Hill, A. Rigosi, Y. Li, O. B. Aslan, D. R. Reichman, M. S. Hybertsen, and T. F. Heinz, Phys. Rev. Lett. **113**, 076802 (2014).

[31]Z. Ye, T. Cao, K. O'Brien, H. Zhu, X. Yin, Y. Wang, S. G. Louie, and X. Zhang, Nature **513**, 214 (2014).

[32]A. Esser, E. Runge, and R. Zimmermann, Phys. Rev. B **62**, 8232 (2000).

[33]P. Kossacki, V. Ciulin, M. Kutrowski, J.-D.Ganie, T. Wojtowicz, and B. Deveaud, Phys. Stat. Sol. (b) **229**, 659 (2002).

[34]R. Xu, J. Yang, Y. Zhu, Y. Han, J. Pei, Y. W. Myint, S. Zhang, and Y. Lu, Nanoscale **8**, 129 (2016).

[35]A.T. Hanbicki, G. Kioseoglou, M. Currie, C. Stephen Hellberg, K.M. McCreary, A.L. Friedman, and B.T. Jonker, Scientific Reports **6**, 18885 (2016).

[36]J. O. Island, G. A. Steele, H. S. J. van der Zant, A. Castellanos-Gomez, 2D Mater. **2**, 011002 (2015).

[37]W. Zhao, Z. Ghorannevis, L. Chu, M. Toh, C. Kloc, P.-H.Tan, and G. Eda, ACS Nano.**7**, 791 (2013).

[38]M. Currie, A. T. Hanbicki, G. Kioseoglou, and B. T. Jonker, Appl. Phys. Lett. **106**, 201907 (2015).

[39]P. K. Chow, R. B. Jacobs-Gedrim, J. Gao, T-M Lu, B. Yu, H. Terrones, and N. Koratkar, ACS Nano **9**, 1520 (2015).

[40]I. Pelant and J. Valenta, Oxford Scholarship, **180** (2012).

[41]Z. Lin, B. R Carvalho, E. Kahn, R. Lv, R. Rao, H. Terrones, Marcos A Pimenta and M. Terrones, 2D Mater. **3**, 022002 (2016).

[42]S. Sasaki, Y. Kobayashi, Z. Liu, K. Suenaga, Y. Maniwa, Y. Miyauchi, and Y. Miyata, Appl. Phys. Express **9**, 071201 (2016).

[43]P. Tonndorf, R. Schmidt, R. Schneider, J. Kern, M. Buscema, G. A. Steele, A. Castellanos-Gomez, H. S. J. van der Zant, S. Michaelis de Vasconcellos, and R. Bratschitsch, Optica **2**, 347 (2015).

[44]P. Hohenberg and W. Kohn, Phys. Rev. B **136**, 864 (1964).

[45]W. Kohn, L. J. Sham, Phys. Rev. A, **140**, 1133 (1965).

[46]K. Schwarz, P. Blaha, G. K. H. Madsen, Comput. Phys. Commun.**147**, 71 (2002).

[47]P. Blaha, K. Schwarz, P. Sorantin, S. B. Trickey, Comput. Phys. Commun. **59**, 399 (1990).

[48]E. Engel, and S. H. Vosko, Phys. Rev. B **50**, 13164 (1993).

[49]A. Zunger and A.J. Freeman. Phys. Rev. B **15**, 5049 (1977).

[50]J. P. Perdew, K. Burke, and M. Ernzerhof, Phys. Rev. Lett. **77**, 3865 (1996).

[51]P. E. Blöchl, O. Jepsen, and O. K. Andersen, Phys. Rev. B **49**, 16223 (1994).

[52]A. H. Reshak , K. Nouneh, I.V. Kityk, J. Bila, S. Auluck, H. Kamarudin, Z. Sekkat, Int. J. Electrochem. Sci. **9**, 955 (2014).

[53]P.Y. Yu and M. Cardona, Springer, Berlin, (1999).

[54]A. Kumar and P. Ahluwalia, Eur. Phys. J. B **85**, 1 (2012).

[55]A. Kuc, N. Zibouche, and T. F. Heinz, Phys. Rev. B **83**, 245213 (2011).

[56]Y. Ding, Y. Wang, J. Ni, L. Shi, S. Shi, and W. Tang, Physica B **406**, 2254 (2011).

[57]L. V. Keldysh, JETP Lett. **29**, 658 (1979).

[58]D. Sanvitto, D. M. Whittaker, A. J. Shields, M. Y. Simmons, D. A. Ritchie, and M. Pepper, Phys. Rev. Lett. **89**, 246805 (2002).

[59]R. A. Sergeev, and R. A. Suris, Nanotechnol. **12**, 597 (2001).

[60]C. H. Lui, A. J. Frenzel, D. V. Pilon, Y.-H. Lee, X. Ling, G. M. Akselrod, J. Kong, and N. Gedik, Phys. Rev. Lett. **113**, 166801 (2014).

[61]F. Gao, Y. Gong, M. Titze, R. Almeida, Pulickel M. Ajayan, and H. Li arXiv:1604.04190.

[62]D. Lagarde, L. Bouet, X. Marie, C. R. Zhu, B. L. Liu, T. Amand, P. H. Tan, and B. Urbaszek, Phys. Rev. Lett. **112**, 047401 (2014).

[63]H. Shi, R. Yan, S. Bertolazzi, J. Brivio, B. Gao, A. Kis, D. Jena, H. G. Xing, and L. Huang, ACS Nano **7**, 1072 (2013).

[64]Q. Wang, S. Ge, X. Li, J. Qiu, Y. Ji, J. Feng, and D. Sun, ACS Nano **7**, 11087 (2013).

[65]H. Wang, C. Zhang, and F. Rana, Nano Lett. **15**, 339 (2015).

[66]X. Zhang, Y. You, S. Y. F. Zhao, and T. F. Heinz, Phys. Rev. Lett. **115**, 257403 (2015).

[67]A. Singh, G. Moody, K. Tran, M. E. Scott, V. Overbeck, G. Berghäuser, J. Schaibley, E. J. Seifert, D. Pleskot, N. M. Gabor, J. Yan, D. G. Mandrus, M. Richter, E. Malic, X. Xu, and X. Li, Phys. Rev. B **93**, 41401 (2016).

[68]L. Yang, W. Chen, K. M. McCreary, B. T. Jonker, J. Lou, and S. A. Crooker, Nano Lett. **15**, 8250 (2015).

[69]L. Yang, N. A. Sinitsyn, W. Chen, J. Yuan, J. Zhang, J. Lou, and S. A. Crooker, Nat. Phys. **11**, 830 (2015).

[70]G. Moody, C. K. Dass, K. Hao, C.-H. Chen, L.-J. Li, A. Singh, K. Tran, G. Clark, X. Xu, G. Berghauser, E. Malic, A. Knorr, and X. Li, Nat. Commun. **6**, 8315 (2015).

[71]A. Srivastava, M. Sidler, A. V. Allain, D. S. Lembke, A. Kis and A. Imamoğlu, Nat. Nanotechnol. **10**, 491 (2015).

[72]M. Koperski, K. Nogajewski, A. Arora, V. Cherkez, P. Mallet, J.-Y. Veuillen, J. Marcus, P. Kossacki, and M. Potemski, Nat. Nanotechnol. **10**, 503 (2015).

[73]B. Chen, H. Sahin, A. Suslu, L. Ding, M. I. Bertoni, F. M. Peeters, and S. Tongay, ACS Nano. **9**, 5326 (2015).

[74]J. Huang, T. B. Hoang, and M. H. Mikkelsen, Scientific Reports **6**, 22414 (2016).

[75]X. Wen-Fang, Chin. Phys. Lett. **22**, 1768 (2005).

[76]V. Savona, J. Phys.: Condens. Matter **19**, 295208 (2007).

[77]S. Jaziri and R. Ferriera, Phys. Stat. Sol. **221**, 337 (2000).

[78]V. Savona, Phys. Rev. B **74**, 075311 (2006).

[79]E. Bernhardt, Stability of rays traveling through Gaussian random potential (2009).

[80]F. A. Rasmussen and K. S. Thygesen, J. Phys. Chem. C **119**, 13169 (2015).

**Table**: The calculated band gap, dielectric constant $\varepsilon_r = \varepsilon_1^\perp$ and effective electron, hole and reduced masses of monolayer WS$_2$ compared to other measured values.

|  | $E_g$ (eV) | $\varepsilon_r$ | $m_e(m_0)$ | $m_h(m_0)$ | $\mu(m_0)$ |
|---|---|---|---|---|---|
| Our work | 2.24 | 5.8 | 0.34 | 0.34 | 0.17 |
| 30 | 2.36 |  |  |  | 0.16 |
| 56 | 2.91 |  |  |  |  |
| 80 | 2.18 |  | 0.46 | 0.42 | 0.22 |

**Figures captions**

**Fig. 1: a)** Band structure and densities of states of WS$_2$ monolayer using DFT approach. The energy at the valence band maximum is set to zero and the gap is direct at the K point. **b)** Real part of parallel and perpendicular dielectric constant ($\varepsilon_1^{//}$ and $\varepsilon_1^\perp$) as a function of photon energy of WS$_2$ monolayer.

**Fig. 2**: **a)** Exciton peak binding energies obtained from the derivative of the reflectance contrast (blue circle)[30] and two-photon absorption (red star)[31] are compared with our theoretical model (black squares). The monolayer WS$_2$ is deposited on the SiO$_2$/Si substrate with $\varepsilon_2 = 3.9\,\varepsilon_0$ and exposed to the air with $\varepsilon_1 = 1$. The dielectric constant, reduced effective mass and the band gap of the monolayer WS$_2$ are given in the table. In the inset, a plot of the 2D radial probability density $\rho|\psi_{1s}(\rho)|^2$ where $\psi_{1s}(\rho)$ is the *1s* exciton wave function which is computed by numerically solving the exciton Hamiltonian $H_X$, including the non-local dielectric screening potential $V_{2D}(\rho)$. Such 2D potential affects the low-energy states because of their small radii and results in a nonhydrogenic Rydberg series. Degenerate states are connected by horizontal lines.

**b)** Screening dielectric potential (filled circles) and Coulomb interaction (open circles) as a function of the dielectric constant of the environment $\varepsilon_2 = 1$ (red), 4 (black) and 10 (blue). Vertical dashed lines represent the screening length $r_s$, showing that the short range interaction is more strongly affected by the dielectric environment. This 2D potential is valid only for $r_s > a_B$ or $\frac{\varepsilon_1+\varepsilon_2}{2} < \varepsilon_r$.

**Fig. 3**: **a)** Exciton and trion PL emission lineshape at 4K. **b)** Temperature dependence of the trion PL emission lineshape.

**Fig. 4**: Mass action law plot for fixed quantity of photons $n^P = 2.10^{10} cm^{-2}$. Trion (open circles) and exciton (filled circles) density as a function of **a)** temperature for *s=1* and **b)** pump to background ratio $s = n^p/n^B$ at *T=100K*. **c)** Trion and **d)** exciton as a function of temperature and *s*.

**Fig. 5:** Temperature dependence of the rate $n_X/n_T$ for different values of pump to background density **a)** *s=0.5* **b)** *s=1* at constant quantity of absorbed photons of $n^P = 2.10^{10} cm^{-2}$; and for several **c)** $n^P = 0.15\ 10^{10} cm^{-2}$ for *s=1* **d)** $n^P = 10^8 cm^{-2}$ for *s=2.5*. The experimental values obtained from micro PL spectra[16] measured for different temperatures are shown as cross.

**Fig. 6:** Localized exciton states and optical spectrum of the WS$_2$ monolayer. **a)** The Gaussian well containing localized excitons is embedded into the dielectric medium. The height and the well radius of the Gaussian potential have been adjusted to give the best agreement with experiment. **b)** PL spectrum of monolayer WS$_2$, the emission band is located at 1860-1815meV, it is about tens of meV below the free exciton peak. For comparison, the PL spectrum of monolayer WSe$_2$ is shown in the inset. The parameters used in these spectra are provided by our ab-initio DFT method except for WSe$_2$ plotted in the inset.

**Fig. 7:** The exciton states subject to 2D static disorder potential for monolayer WS$_2$ sample. **a)** Spatial map of disorder potential showing lakes-and-hills inhomogenous landscape. **b)** Low-temperature (10K) PL spectra of a monolayer WS$_2$ flake. The inhomogenously broadened exciton resonance (the defect band) is located around 100 meV below the neutral free A-exciton peak. This computed states are in agreement with experimental measurements.[39] **c-e)** Probability distribution $\left|\Psi_{\tilde{n}_X,\tilde{n}_Y,1s}(R,\rho,\theta)\right|^2$ of selected excitonic states as labeled. The colour scale is the normalized wavefunction probability and applies to panels c-e). The simulation parameters used here are the exciton correlation length $L = 45\text{Å}$ and the potential depth $V_0 = 110 meV$.

**Fig. 1a**

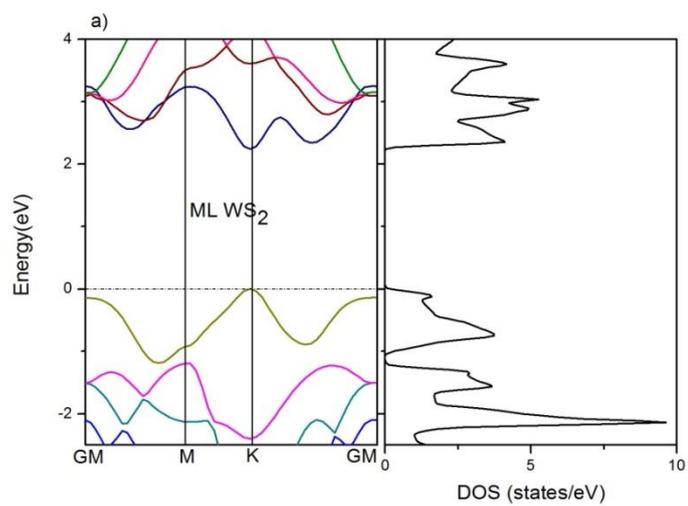

**Fig. 1b**

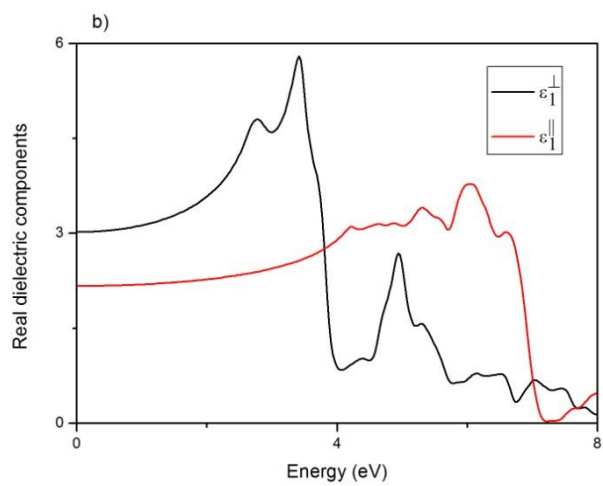

**Fig. 2a**

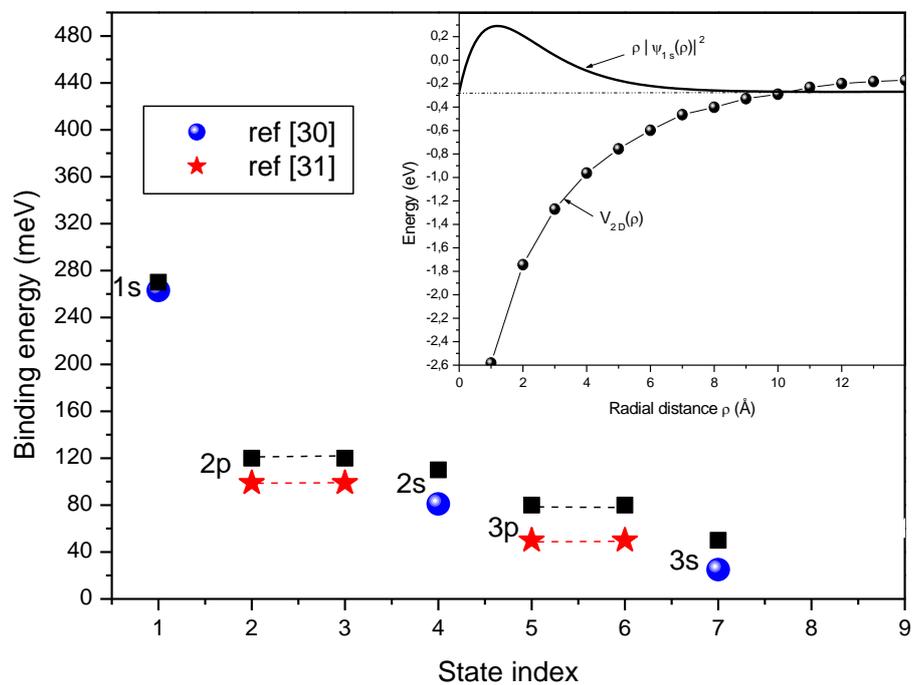

**Fig. 2b**

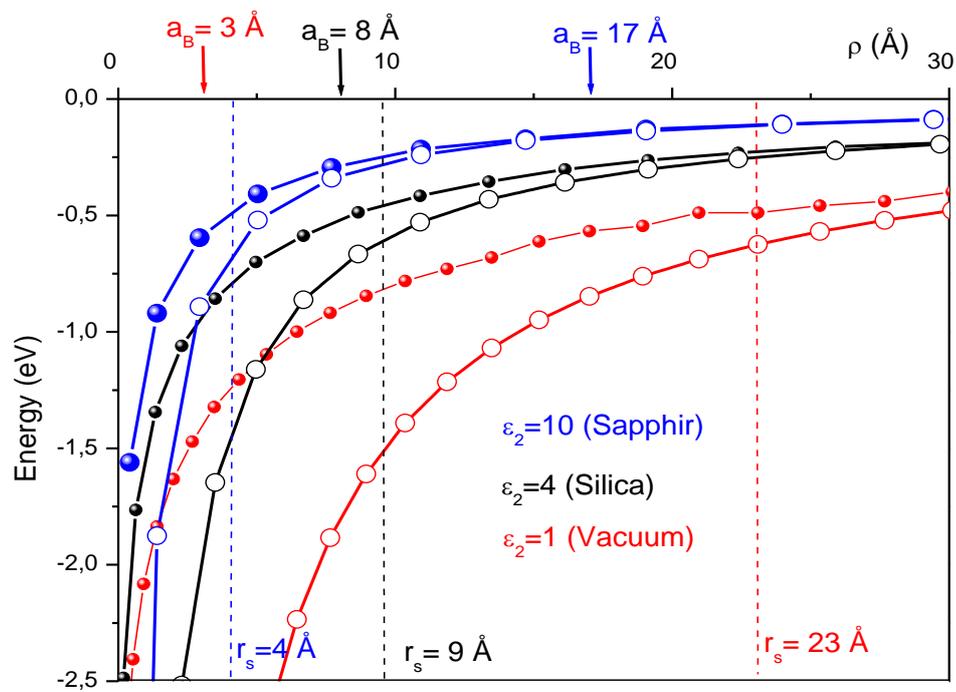

**Fig.3a**

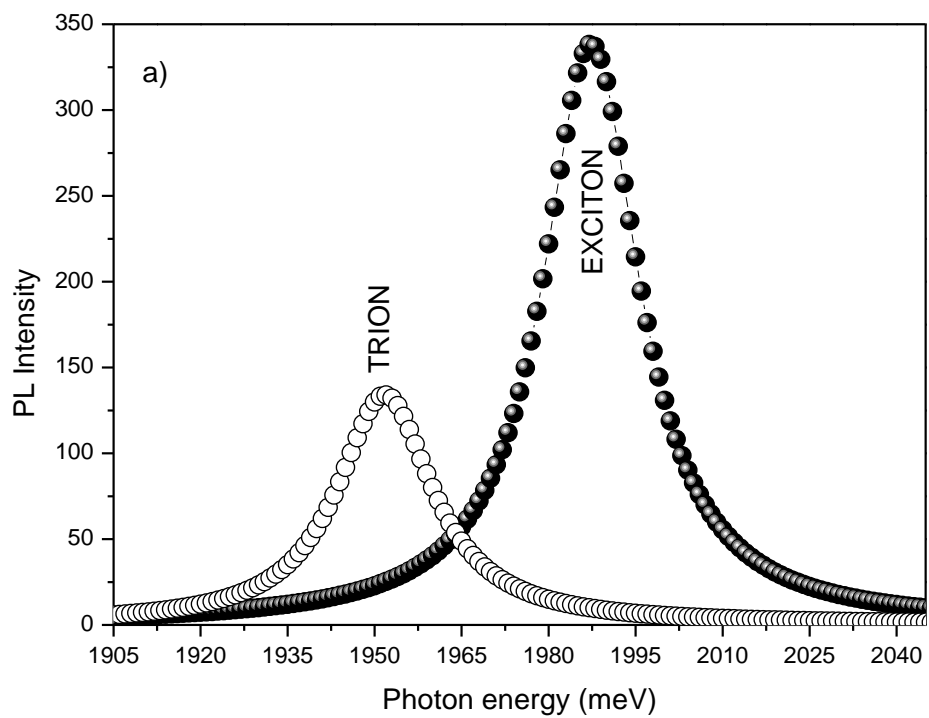

**Fig.3b**

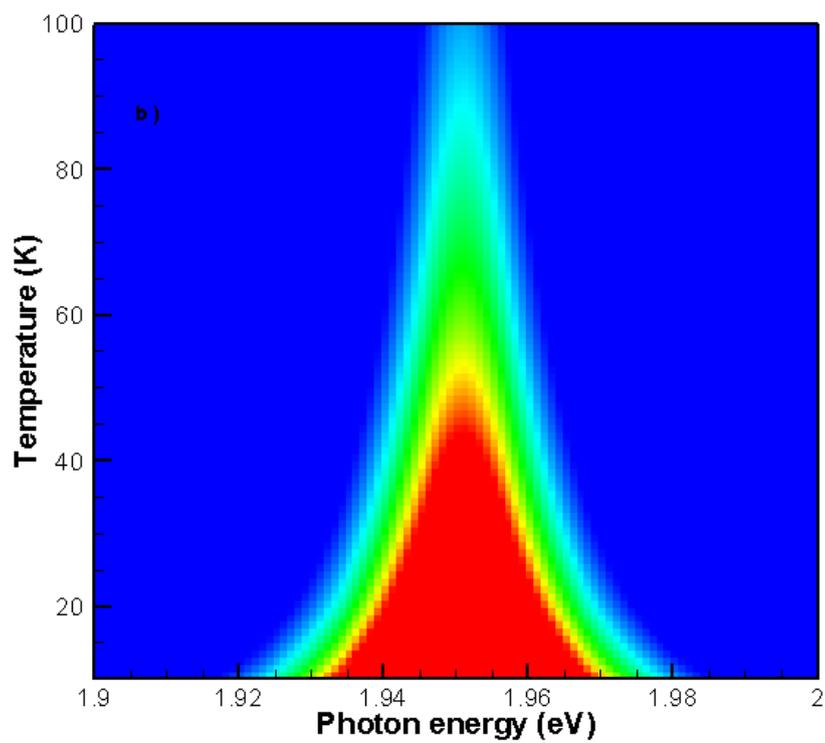

**Fig. 4a**

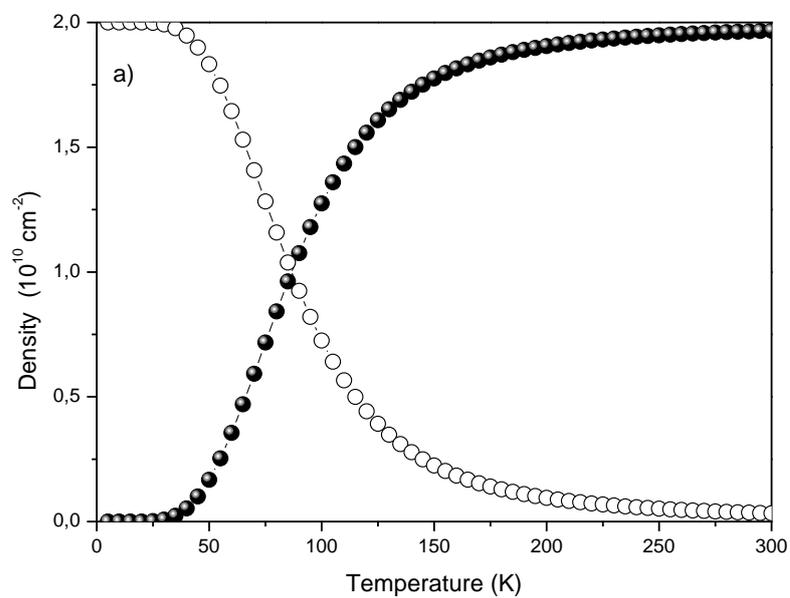

**Fig. 4b**

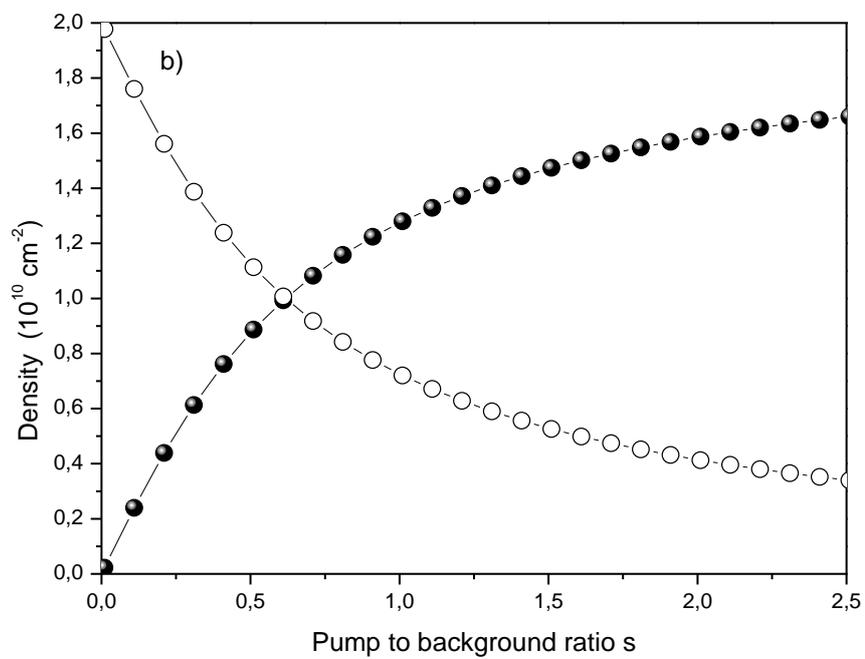

**Fig. 4c**

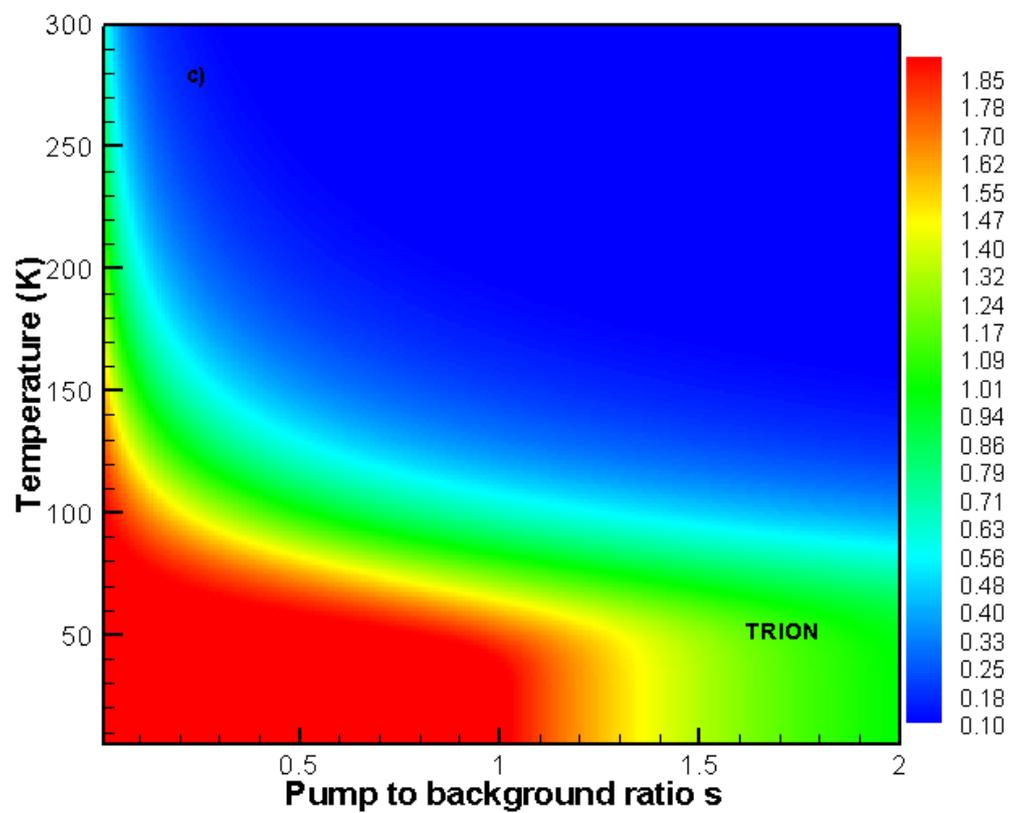

**Fig. 4d**

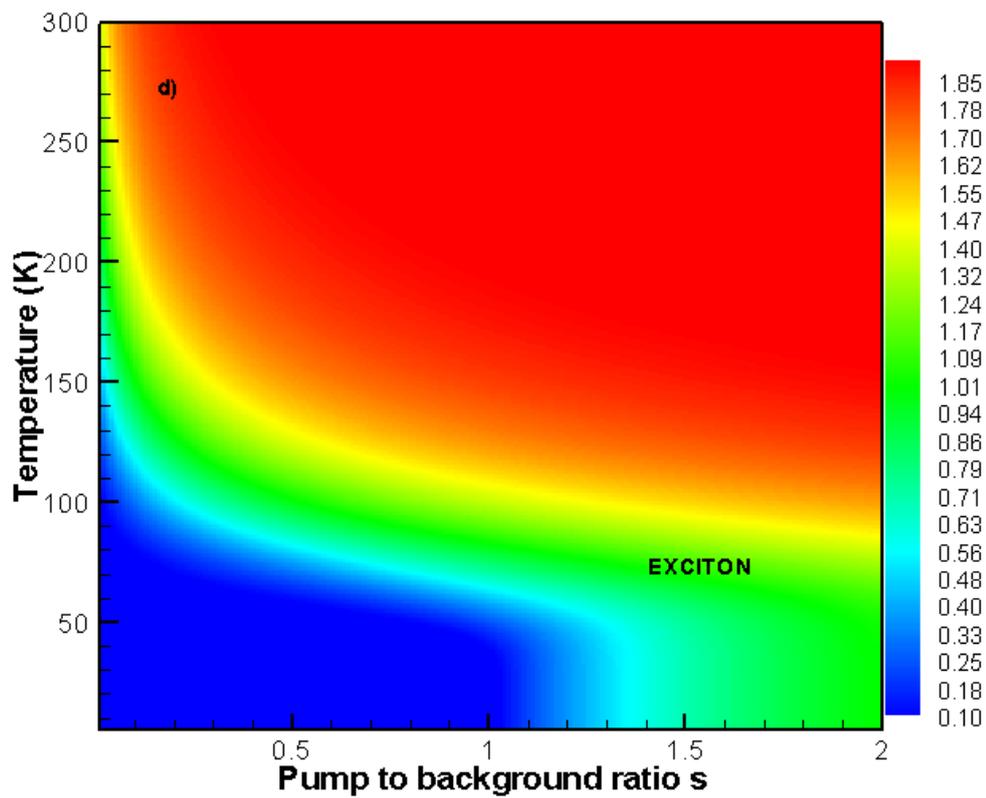

**Fig. 5**

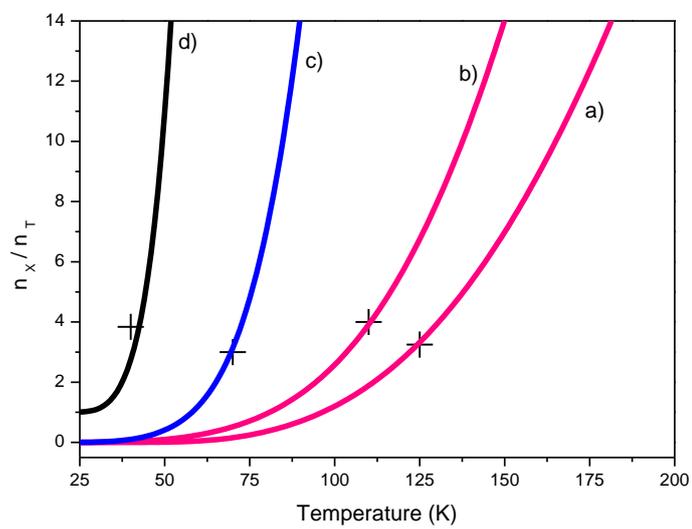

**Fig. 6a**

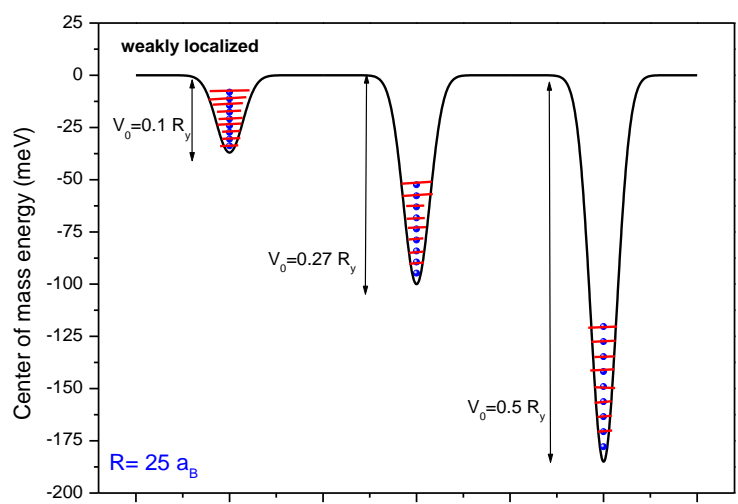

**Fig. 6b**

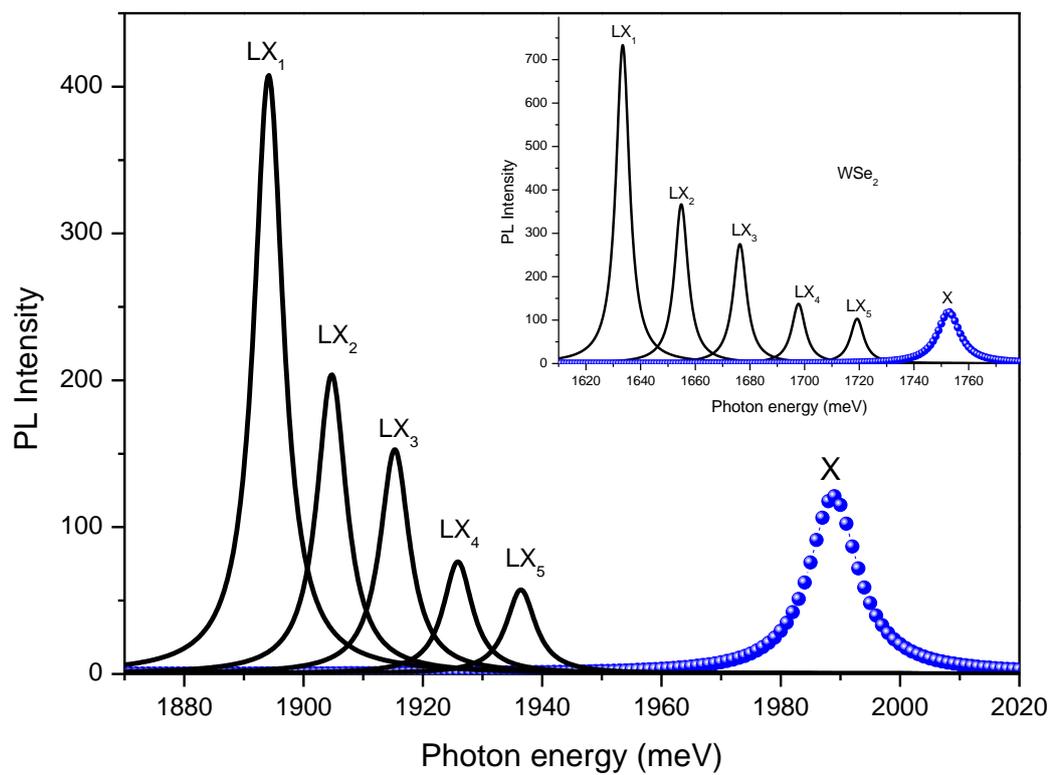

**Fig. 7 a**

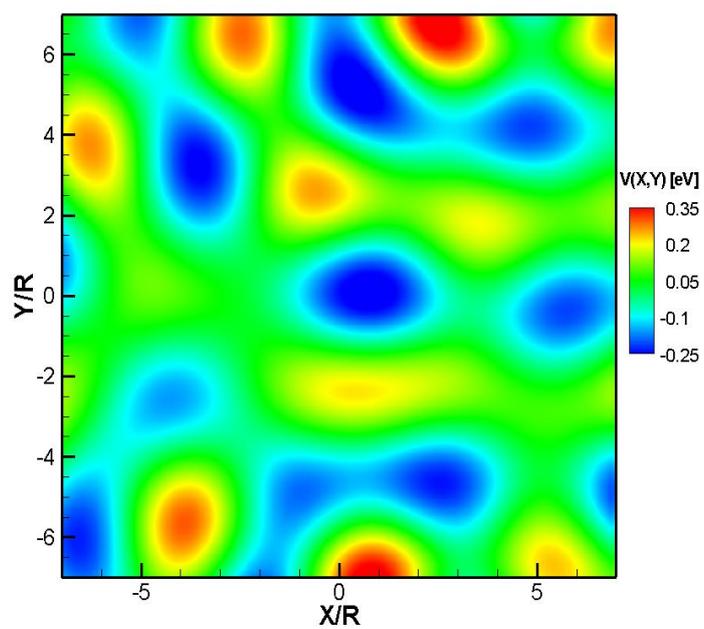

**Fig. 7 b**

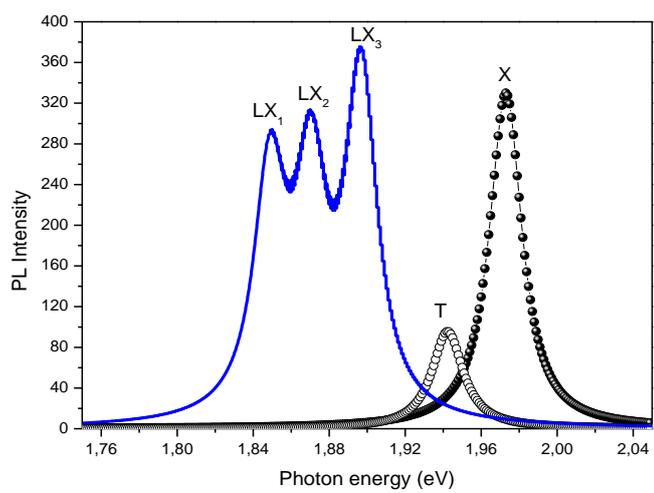

**Fig. 7 c-e**

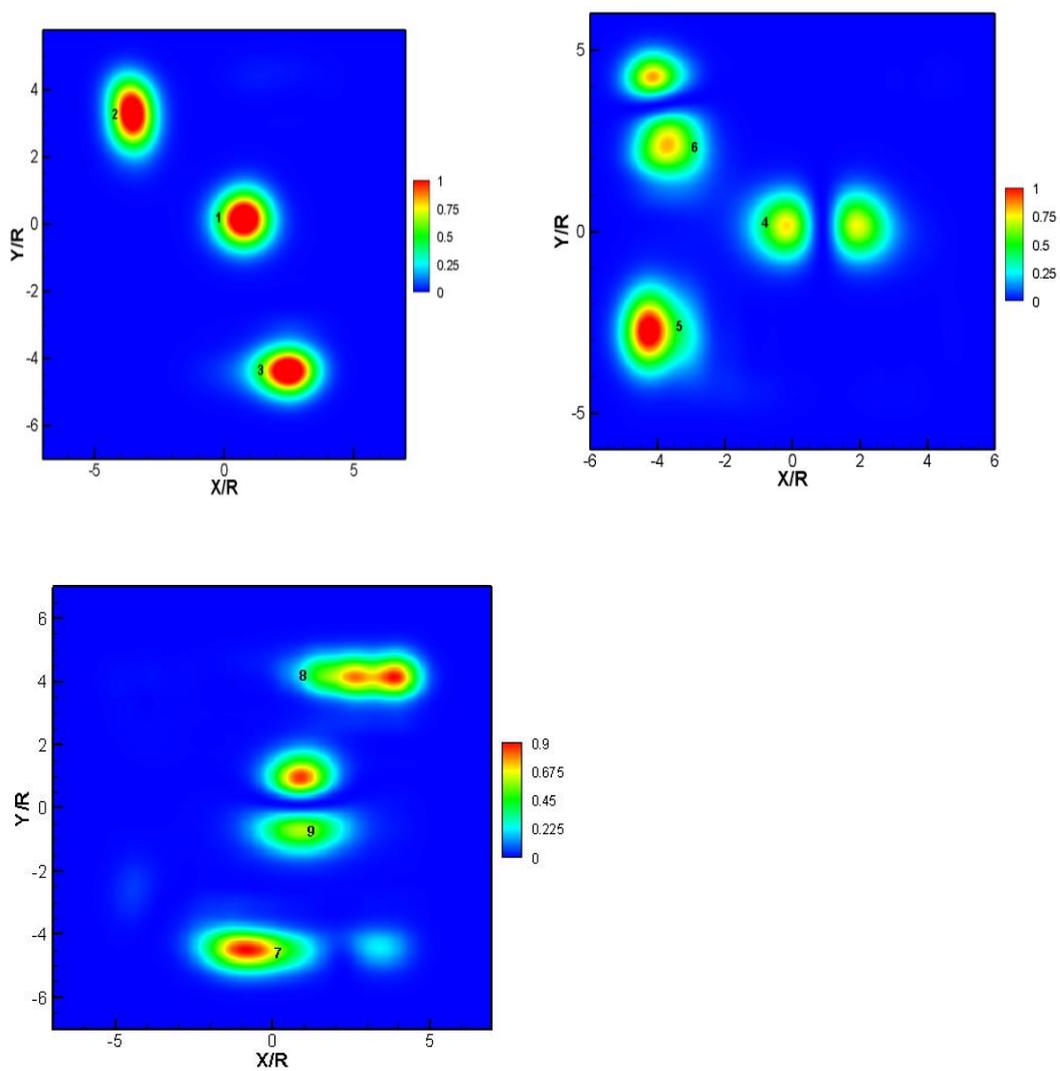